\documentclass[prb,twocolumn,amsmath,amsfonts,showpacs]{revtex4-2}
\usepackage{graphicx}
\usepackage{epsf}
\usepackage{epstopdf}
\usepackage{graphicx}
\usepackage{tikz}
\usepackage{float}
\usepackage{amsmath,bm,upgreek}
\usepackage{nccmath}
\usepackage[mathscr]{euscript}
\usepackage{natbib}
\usepackage{hyperref}

\begin{document}
\draft
\title{Coherent current correlations in a double-dot Cooper pair splitter}
\author{Bogdan R. Bu{\l}ka}
\affiliation{Institute of Molecular Physics, Polish Academy of
Sciences, ul. M. Smoluchowskiego 17, 60-179 Pozna{\'n}, Poland}

\date{Received \today \hspace{5mm} }

\begin{abstract}
Exact analytical formulas are derived, by means of Keldysh Green functions, for currents and current correlation functions in a Cooper pair splitter modelled on a double quantum dot system coherently coupled to a superconductor and two normal metallic electrodes.
Confining to the subspace with the inter-dot singlet we show perfect entanglement of split electrons in two separated crossed Andreev reflection processes. The studies are focused on the noise power spectrum in a whole bias voltage range. In particular, in the large voltage limit shot noise dominates and its spectrum exhibits two extraordinary side dips related to resonant inter-level current correlations caused by coherent electron-hole recombination processes accompanied by emission and absorption of photons.
In the linear response limit we derived the frequency dependent admittance which shows different interference patterns for the cross and the auto current correlations.

\end{abstract}

\maketitle

\section{Introduction}

Electronic transport through a nanoscopic system is stochastic in nature, and therefore, in investigations one needs to measure and analyze the  currents, the second-order current correlation functions (noise power) and all other high-order cumulants to get the full counting statistics (FCS)  \cite{Levitov1992,Levitov1996,Nazarov2003}. In particular, the current-current correlations give insight into quantum noise in nanoscopic electrical circuits~\cite{Clerk}, shot noise and dynamics of charge transfers in the presence of interactions \cite{BlanterButtiker},  entanglement of scattered particles \cite{Burkard2000,Lesovik2001,Recher2001,Samuelsson,Beenakker2006}, their fermionic or bosonic nature~\cite{Jeltes2007} as well as the effective charge in the fractional quantum Hall effect \cite{Saminadayar,Picciotto1997,Reznikov1999}.

Ubbelhode et.  al.~\cite{Ubbelohde2012} used a quantum point contact  as a  highly sensitive counter of charge transport in a single-electron transistor. This setup allowed them to determine directly the current statistics and timescales of the current fluctuations (up to tens of $\mu$s) as well as the frequency-dependent third order correlation function (the skewness).
An active quantum detector system was proposed \cite{Lesovik1997,Gavish,Schoelkopf2003}  to measure quantum noise, and observe zero-point fluctuations. Such a device was fabricated \cite{Aguado} on a double quantum dot as a tunable two-level system and showed current fluctuations in the capacitive coupled  conductor over a very wide frequency range (up to hundreds of GHz).  An alternative quantum detector was based on a superconductor-insulator-superconductor tunnel junction as an on-chip spectrum analyzer \cite{Clerk2002}, which allowed one to measure the nonsymmetrized current noise arising from coherent charge oscillations in a superconducting charge qubit~\cite{Deblock2003,Xue2009}, or  generated by the tunneling of quasiparticles across a Josephson junction polarized
in the vicinity of the superconducting gap \cite{Billangeon2006,Billangeon2009}. A similar experimental setup was used to  investigate the nonequilibrium dynamics of many-body phenomena on the nanoscale, namely to measure the high frequency current fluctuations of a carbon nanotube quantum dot in the Kondo regime \cite{Basset}.

In electronic transport one can distinguish two regimes: sequential tunneling and coherent (ballistic) transport. In the sequential regime following tunneling events are independent and the transport is described within the Markov approach, using the quantum master equation \cite{Gardiner}.  In particular, the current correlation functions can be derived \cite{vliet,Korotkov,Kogan}, one can get easily FCS \cite{Bagrets,Belzig,Nazarov,Schaller} with all order cumulants for the zero-frequency case, as well as, with some effort, frequency-dependent cumulants \cite{Marcos2010}. Using the spectral decomposition of the noise power spectrum one can distinguish various relaxation processes~\cite{vliet}, related to local charge and spin fluctuations, as well as  see inter-channel current correlations, which can lead to the sub- or super-Poissonian type of shot noise (showing anti-bunching or bunching of transferred particles) \cite{Bulka2000,Bulka1999}.

Noise power in the coherent regime was investigated by means of the scattering approach by
B{\"u}ttiker and coworkers (in many seminal papers; see the review \cite{BlanterButtiker}).
The non-equilibrium Green function  technique~\cite{Haug} is very efficient, it can be used to determine zero-frequency cumulants within FCS  \cite{Nazarov1999,Belzig2003,Gogolin2006}.
The spectral analysis of the current correlations plays a key role in electron quantum optics, a new branch of nanoelectronics, where single electron packets (the levitons) can propagate ballistically along the edge channels  of the quantum Hall effect and can be guided in optics like setup \cite{Parmentier}. In particular, the first and second-order correlation functions are used in description of two-particle interferences in the Hanbury-Brown and Twiss (HBT) geometry \cite{Bocquillon,Glattli}  and in the electron analog the Hong, Ou and Mandel (HOM) experiment \cite{Marguerite} (see also \cite{Glattli2017}).

We are interested in quantum coherence processes in the noise power spectrum and the Cooper pair splitter (CPS) seems appropriate for this issue.
Such system consists of three electrodes in a Y geometry, where the central electrode is a superconductor and serves as a reservoir of Cooper pairs which are injected into two normal metallic electrodes.~\cite{Burkard2000,Lesovik2001,Recher2001,Borlin2002,Sauret2004} This is the fermionic analog of two photon interference experiments: HBT and HOM, where spin-entangled electrons, as Einstein-Podolsky-Rosen (EPR) pairs, are spatially coherently separated into their entangled constituents.
The device allows one to test Bell inequalities in terms of current-current correlations  and to show their violation as evidence of entanglement of electrons~\cite{Chtchelkatchev2002, Samuelsson2003,Busz2017}.
Therefore, the CPS is considered as the solid state setup for quantum communications~\cite{Wendin2017}.

The efficiency of CPS can increase when the superconductor is coupled by two quantum dots (QD) with two normal metallic  electrodes~\cite{Recher2001,Sauret2004}.
If the intra-dot charging energy is large, then two electrons of the Cooper pair are forced to split into separate channels. Such devices with tunable QDs were fabricated and shown to control the Cooper pair splitting with a very high efficiency~\cite{Hofstetter2009,Herrmann2010,Hofstetter2011, Schindele2012, Das2012}; recently also with graphene QDs \cite{Tan2015} and two topologically non-trivial semiconducting nanowires \cite{Baba2018}.

Chevallier at al. ~\cite{Chevallier} modeled the double-dot Cooper pair splitter (2QD-CPS) and calculated the current as well as the current correlations by means of non-equilibrium Green functions. Cross Andreev reflections (CAR)  are relevant for operation of CPS. In the CAR process an electron ejected from the normal metallic  electrode to the superconductor forms a Cooper pair and simultaneously a hole of opposite spin is injected to the other metallic  electrode. Their excitation energies are less than the superconducting energy gap.
There are two other processes: direct Andreev reflection (DAR), when the electron and the hole propagate to the same electrode, and electron  cotunneling (EC) related to electron transfers from one metallic electrode  to the other one through the superconductor. However, these processes spoil the splitting efficiency.
The anomalous electron-hole current correlations (for CAR and DAR) are positive~\cite{Anantram1996}, which is related to bunching of tunneling events~\cite{Burkard2000}. This is in contrast to the EC processes which result in negative cross correlations~\cite{Anantram1996}, as in the normal metallic Y splitter~\cite{BlanterButtiker,Bulka2008}. The interplay between these processes was demonstrated in the 2QD-CPS model with two ferromagnetic electrodes~\cite{Wrzesniewski2017}.

The above mentioned studies
 focused solely on the zero-frequency coherent current correlations. Our aim is to study the noise power spectrum  in 2QD-CPS to get insight into quantum coherence processes in transport, a relevant time scale for Cooper pair splitting, local charge dynamics and relaxation processes.
 Recently, Droste et al.~\cite{Droste} studied the frequency dependent noise power in a two terminal hybrid system, composed of a single-level quantum dot coherently coupled to a superconductor and a normal conducting electrode (N-QD-S). They found that the noise power spectrum reflects the internal spectrum of the proximized dot and shows extraordinary resonance dips at frequencies corresponding to transitions between the Andreev bound state (ABS).
Our goal is to investigate a role of quantum interference in the second-order current correlation functions, in particular, the origin of the dips in the noise power spectrum. We predict that interference processes in the noise power exhibit themselves in a different way than Young's interference patterns observed in the first-order correlations.

The paper is organized as follows. In Sec. \ref{model} we present the model and the Keldysh Green function method, following  Chevallier at al. ~\cite{Chevallier} with some modifications which make the presentation more transparent. The considerations are focused on the subgap energy region with the ABS reflections only.
In Sec. \ref{currents} we derive analytical formulas for the current and the frequency dependent (cross and auto) current correlation functions.   The analysis of the results is presented in Sec. \ref{results}: first the zero-frequency case  (Sec.\ref{zero-frequency}) and next the main results of the paper on the frequency dependence of the noise power (Sec.\ref{frequency}). The results are presented  for a whole range of the bias voltage: the shot noise spectrum in the large voltage limit, down to the linear response limit with the frequency dependent admittance. Supplemental materials are placed in Appendixes \ref{appA} and \ref{appB}.

\section{Model description and Green function calculations}
\label{model}

\begin{figure}
\centering
\includegraphics[width=0.5\linewidth,clip]{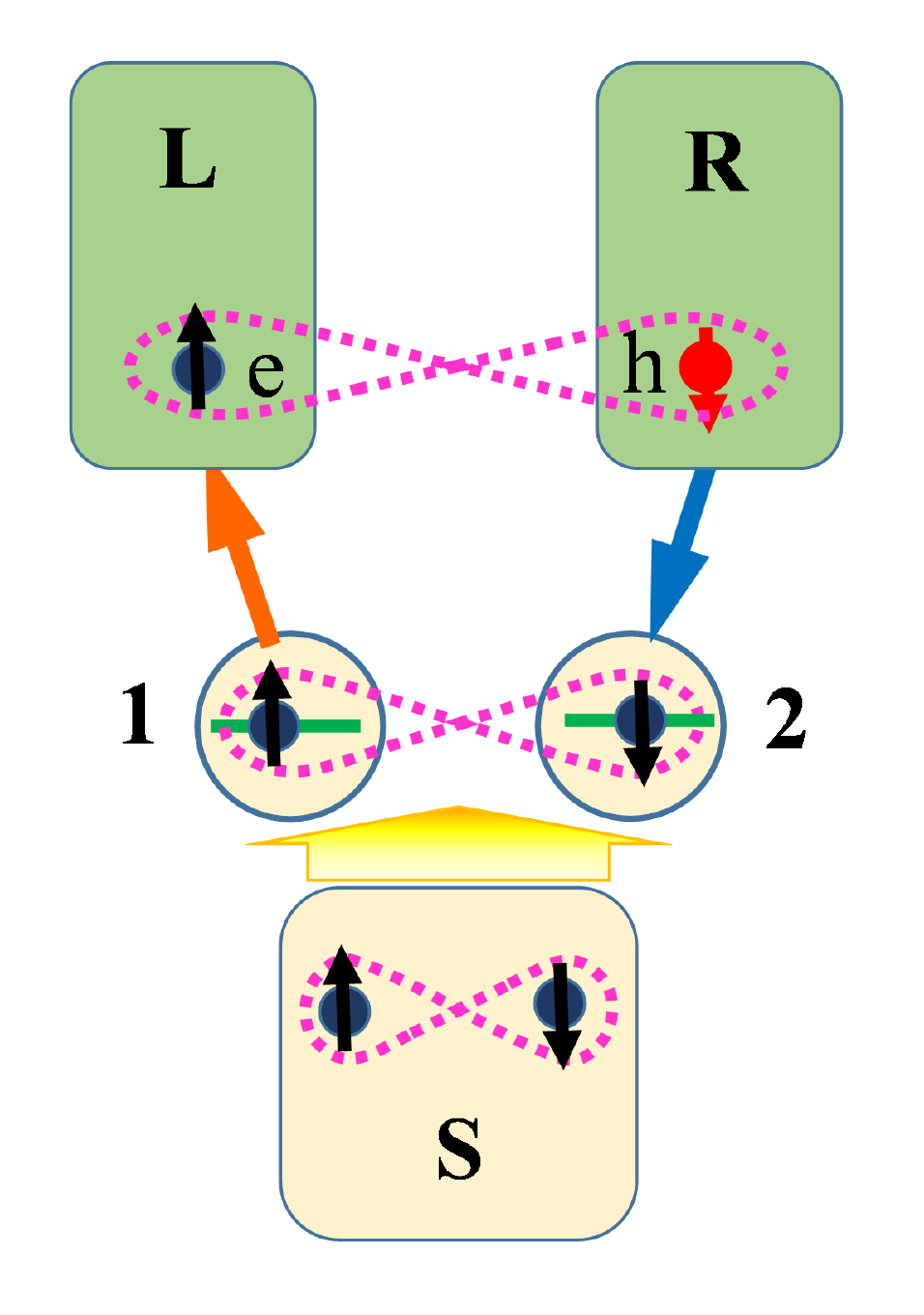}
\caption{Schematic presentation of the Cooper pair splitter with two quantum dots (1,2) coupled to the normal ($L, R$) electrodes and strongly coupled to the superconductor ($S$) as a reservoir of Cooper pairs. Transport is due to perfect cross Andreev reflections  when an electron ($e$) is injected to the normal electrode and a hole ($h$) with an opposite spin is simultaneously ejected from the second metallic electrode. }
\label{fig1}
\end{figure}

We consider a Cooper pair splitter with two quantum dots (2QD), where each QD is coupled to the normal $L$ or $R$ electrode and both are coupled the superconducting BCS reservoir of Cooper pairs; see Fig.1.
Moreover, we assume the case of strong coupling to the $S$ electrode when due to proximity effect the QDs behave like superconducting grains and entangled electrons can be transported through the QDs.
The model is described by the Hamiltonian
\begin{equation}
\label{eq:ham}
H = \sum_{i=1,2;\sigma=\uparrow,\downarrow}\epsilon_{i} a^{\dag}_{i\sigma}a_{i\sigma} + \sum_{\alpha=L,R,S}H_{\alpha} + H_{T} \; ,
\end{equation}
where the first term
corresponds to the 2QD system with a single level $\epsilon_{i}$ available at  the $i$-th QD.
The electrode Hamiltonians
\begin{eqnarray}
H_{\alpha}=\sum_{k}\Psi^{\dag}_{\alpha k}(\varepsilon_{\alpha k}\sigma_z+\Delta_{\alpha}\sigma_x)\Psi_{\alpha k},
\end{eqnarray}
are expressed in Nambu notation $\Psi^{\dag}_{\alpha k}=[c^{\dag}_{\alpha k\uparrow},c_{\alpha \bar{k}\downarrow}]$, $\bar{k}=-k$,
$\sigma_z$, $\sigma_x$ are the Pauli matrices, $\varepsilon_{\alpha k}$ and $\Delta_{\alpha}$ denote the electron energy and the superconducting gap, which in the normal electrodes is taken as zero.
The transfer Hamiltonian is given by
\begin{eqnarray}\label{eq:transferH}
H_T=\sum_{k}\big(\Psi^{\dag}_{L k}t_{L1}\sigma_z d_{1}+\Psi^{\dag}_{R k}t_{R2}\sigma_z d_{2}+\nonumber\\
\Psi^{\dag}_{S k}t_{S1}\sigma_z d_{1}+\Psi^{\dag}_{S k}t_{S2}\sigma_z d_{2}+h.c.\big),
\end{eqnarray}
where $d^{\dag}_{i}=[a^{\dag}_{i\uparrow},a_{i\downarrow}]$ and $t_{\alpha i}$ describes electron hopping between the $\alpha$ electrode and the $i$-th QD.

 The charge current operator from the $\alpha$ electrode to the $i$-th dot is
\begin{eqnarray}\label{eq:currentop}
\hat{I}_{\alpha i} \equiv\left(
\begin{array}{l}
\hat{I}_{\alpha i e}\\
\hat{I}_{\alpha i h}
\end{array}
\right)=\imath e \sum_{\alpha,k}(\Psi^{\dag}_{\alpha k}t_{\alpha i}\sigma_z d_{i}-d^{\dag}_{i}t_{i\alpha}\sigma_z\Psi_{\alpha k}),
\end{eqnarray}
where $\hat{I}_{\alpha i e}$ and $\hat{I}_{\alpha i h}$ denote  contributions by electrons and holes, respectively.

To calculate the average of the currents and their correlation functions we use the Keldysh Green function method~\cite{Chevallier,Dong,Zazunov,Cuevas}. Since the considered model is for noninteracting particles one can easily write the equation of motion for the Keldysh Green functions. In derivations the interdot singlet, with $\langle a^{\dag}_{1\uparrow}a^{\dag}_{2\downarrow}-a^{\dag}_{1\downarrow}a^{\dag}_{2\uparrow}\rangle \neq 0$, is only taken into account.
We assume that intra-dot Coulomb interactions are large and double occupancy of the dots is neglected,  $\langle a^{\dag}_{i\uparrow}a^{\dag}_{i\downarrow}\rangle = 0$; thus, the transfer of the Cooper pair through a single QD as a high energy process is forbidden. In the first step the self-energy $\hat{\Sigma}_S$ (in the Keldysh-Nambu space) describing the Cooper pair transfer from the S-electrode is derived according to Eq.(26) in Ref.[\onlinecite{Chevallier}] -- see Appendix \ref{appA} for more details.
Our derivations are confined to the subgap regime $|E|<\Delta$  and in the limit $\Delta \rightarrow \infty$ in which  the self-energy becomes
\begin{eqnarray}\label{eq:S-S}
\hat{\Sigma}_{S}=\frac{\gamma_{S}}{2}\left[\begin{array}{cccc}
0&0&1&0\\
0&0&0&-1\\
1&
 0&0&0\\
 0&-1&0&0
\end{array}\right],
\end{eqnarray}
where $\gamma_{S}=\pi \rho_S t_{S1}t_{S2}$ and $\rho_S$ is the density of states in the S-electrode in the normal state. Notice that $\gamma_{S}$  describes the exchange electron-hole coupling between the quantum dots. In this way the S-electrode is integrated out and the Keldysh Green function matrix, for the system with two normal electrodes and two proximized QDs, is expressed as a product of two components
\begin{eqnarray}\label{eq:greentot}
\hat{G}_{L2QDR}=\hat{G}_{e\uparrow,h\downarrow}\otimes \hat{G}_{h\downarrow,e\uparrow},
\end{eqnarray}
where
\begin{widetext}
\begin{gather}
\hat{G}_{e\uparrow,h\downarrow}\equiv  \left[\begin{array}{cccc}
  \hat{G}_{Le\uparrow,Le\uparrow} & \hat{G}_{Le\uparrow,1e\uparrow} & \hat{G}_{Le\uparrow,2h\downarrow}&\hat{G}_{Le\uparrow,Rh\downarrow} \\
  \hat{G}_{1e\uparrow,Le\uparrow} & \hat{G}_{1e\uparrow,1e\uparrow} & \hat{G}_{1e\uparrow,2h\downarrow}&\hat{G}_{1e\uparrow,Rh\downarrow} \\
 \hat{G}_{2h\downarrow,Le\uparrow} & \hat{G}_{2h\downarrow,1e\uparrow} & \hat{G}_{2h\downarrow,2h\downarrow}&\hat{G}_{2h\downarrow,Rh\downarrow} \\
  \hat{G}_{Rh\downarrow,Le\uparrow} & \hat{G}_{Rh\downarrow,1e\uparrow} & \hat{G}_{Rh\downarrow,2h\downarrow}&\hat{G}_{Rh\downarrow,Rh\downarrow}
\end{array}\right]
=\left[\begin{array}{cccccccc}
  w_{L,11}^{--} & w_{L,11}^{-+} & t_{L1}&0 &0 &0 &0&0 \\
  w_{L,11}^{+-} &  w_{L,11}^{++} &  0&-t_{L1} &0 &0 &0&0 \\
t_{L1}& 0 & z_{1e}& 0&\gamma_{S}/2 &0 &0 &0 \\
   0 &  -t_{L1}& 0& -z_{1e}&0 &-\gamma_{S}/2 &0 &0 \\
0 &0 &\gamma_{S}/2 &0 &  z_{2h} & 0 & -t_{R2}&0   \\
0 &0 &0 &-\gamma_{S}/2 & 0 &   -z_{2h} & 0 & t_{R1}  \\
0 &0 &0 &0 &  - t_{R2}& 0 & w_{R,22}^{--}& w_{R,22}^{-+}  \\
0 &0 &0 &0 &    0 & t_{R2}&   w_{R,22}^{+-} & w_{R,22}^{++}
\end{array}\right]^{-1}.\label{eq:NEGF}
\end{gather}
\end{widetext}

Here, we use the Keldysh-Nambu matrix notation. For example the Green function for the $L$ electrode decoupled from the system is expressed as
\begin{gather}
\hat{g}_{Le\uparrow,Le\uparrow}=
\left[\begin{array}{cc}
  g_{Le\uparrow,Le\uparrow}^{--} &   g_{Le\uparrow,Le\uparrow}^{-+}   \\
  g_{Le\uparrow,Le\uparrow}^{+-} &   g_{Le\uparrow,Le\uparrow}^{++}
\end{array}\right],
\end{gather}
where its inverse elements derived in the wide-band approximation are: $w_{L,11}^{--}= w_{L,11}^{++}=-2 \imath \rho_L (f_{Le}-1/2)$, $w_{L,11}^{-+}= 2 \imath \rho_L f_{Le}$, $w_{L,11}^{+-}= -2 \imath \rho_L (1-f_{Le})$.  Similarly,  the inverse elements of $\hat{g}_{Rh\downarrow,Rh\downarrow}$ are: $w_{R,22}^{--}= w_{R,22}^{++}=-2 \imath \rho_R (f_{Rh}-1/2)$, $w_{R,22}^{-+}= 2 \imath \rho_R f_{Rh}$, $w_{R,22}^{+-}= -2 \imath \rho_R (1-f_{Rh})$.
$f_{\alpha e}=\{\exp[(E-\mu_{\alpha})/k_BT]+1\}^{-1}$ and  $f_{\alpha h}=\{\exp[(E+\mu_{\alpha})/k_BT]+1\}^{-1}$ are the Fermi distribution functions for electrons and holes in the $\alpha$ electrode with the chemical potential $\mu_{\alpha}$. The chemical potential in the superconductor is taken $\mu_S=0$. We also denoted $z_{1e}=E-\epsilon_1$ and $z_{2h}=E+\epsilon_2$.
Confining to the subgap regime one gets the inter-dot coupling $\gamma_{S}=\pi \rho_S t_{S1}t_{S2}$, where $\rho_S$ is the density of states in the S-electrode in the normal state.

Similarly one can express $\hat{G}_{h\downarrow,e\uparrow}^{-1}$ replacing the indexes $\{L,1,\uparrow\}\leftrightarrow \{R,2,\downarrow\}$ in Eq.(\ref{eq:NEGF}).

\section{Currents and their correlation functions}
\label{currents}

Using the Keldysh Green function (\ref{eq:NEGF}) one can calculate the currents as
\begin{eqnarray}\label{eq:curle}
I_{Le\uparrow} = I_{Rh\downarrow}=-\frac{e}{\hbar}\int \frac{dE}{2\pi}(f_{Le} - f_{Rh}) \mathcal{T}_{Aeh}(E)
,\\
\label{eq:curlh}
I_{Lh\downarrow} = I_{Re\uparrow}=-\frac{e}{\hbar}\int \frac{dE}{2\pi}(f_{Re} - f_{Lh}) \mathcal{T}_{Ahe}(E)
,
\end{eqnarray}
where the transmission probability is expressed as
\begin{eqnarray}
\mathcal{T}_{Aeh}(E)=\frac{16\gamma_{L}\gamma_{R} \gamma_{S}^2}{|4(z_{1e}- \imath \gamma_{L}) (z_{2h}- \imath \gamma_{R})- \gamma_{S}^2|^2},
\label{eq:TAeh}\\
\mathcal{T}_{Ahe}(E)=
\frac{16\gamma_{L}\gamma_{R} \gamma_{S}^2}{|4(z_{1h} - \imath \gamma_{L}) (z_{2e} - \imath \gamma_{R})- \gamma_{S}^2|^2}
\label{eq:TAhe}
\end{eqnarray}
and the couplings are $\gamma_{L}=\pi \rho_{L} |t_{L1}|^2$,  $\gamma_{R}=\pi \rho_{R} |t_{R2}|^2$. It is seen that  $\mathcal{T}_{Aeh}(E)$ and $\mathcal{T}_{Ahe}(E)$ have a resonant shape with two peaks, associated with  the Andreev bound states (ABS), at the poles $E_{\pm}=[\epsilon_1-\epsilon_2\pm\imath (\gamma_L+\gamma_R)\pm \sqrt{(\epsilon_1+\epsilon_2\pm\imath (\gamma_L-\gamma_R))^2+\gamma_S^2}]/2$ and $E_{\pm}=[-\epsilon_1+\epsilon_2\pm\imath (\gamma_L+\gamma_R)\pm \sqrt{(\epsilon_1+\epsilon_2\mp\imath (\gamma_L-\gamma_R))^2+\gamma_S^2}]/2$, respectively.
Although $\mathcal{T}_{Aeh}(E)$ and $\mathcal{T}_{Ahe}(E)$ can be different, they have the relative electron-hole symmetry, and therefore the currents $I_{Le\uparrow}$ and $I_{Lh\downarrow}$  are equal.

To study dynamical correlations of the currents we introduce the operator $\Delta \hat{I}_{\alpha i}(t)\equiv \hat{I}_{\alpha i}(t) -\langle I_\alpha \rangle $
 describing current fluctuation from its average value, and next define the non-symmetrized correlation function (following  \cite{Nazarov})
 \begin{align}
S_{\alpha i,\alpha' i'}(\omega)=2\int_{-\infty}^{\infty}dt\; e^{-i\omega t} \langle\Delta \hat{I}_{\alpha i}(0)\Delta \hat{I}_{\alpha' i'}(t)\rangle
\end{align}
for the current in the junction $\alpha i$ and the current in the junction $\alpha' i'$. To measure quantum noise, one needs a quantum detector \cite{Lesovik1997,Aguado,Clerk}, which can absorb or emit energy from noise, for positive and negative frequencies, respectively.

First we derive the cross current correlation function $S_{Le\uparrow,Rh\downarrow}$. Using the current operators, Eq.(\ref{eq:currentop}), one gets the function as a sum of two particle averages, which are decoupled by means of Wick's theorem to products of single particle averages, and next we express them by the Green functions. The result is
\begin{widetext}
\begin{align}
S_{Le\uparrow,Rh\downarrow}(\omega)
=\frac{2e^2}{\hbar}t_{L1}t_{R2}\int_{-\infty}^{\infty} \frac{dE}{2\pi}
\big[ \;G^{+-}_{Rh\downarrow,Le\uparrow}(E)G^{-+}_{1e\uparrow,2h\downarrow}(E+\hbar\omega)
-G^{+-}_{2h\downarrow,Le\uparrow}(E)G^{-+}_{1e\uparrow,Rh\downarrow}(E+\hbar\omega)\nonumber\\
-G^{+-}_{Rh\downarrow,1e\uparrow}(E)G^{-+}_{Le\uparrow,2h\downarrow}(E+\hbar\omega)
+G^{+-}_{2h\downarrow,1e\uparrow}(E)G^{-+}_{Le\uparrow,Rh\downarrow}(E+\hbar\omega)\big] .
\end{align}
Similarly, the auto-correlation function is expressed as
\begin{eqnarray}
S_{Le\uparrow,Le\uparrow}(\omega)=\frac{2e^2}{\hbar}t_{L1}^2\int_{-\infty}^{\infty} \frac{dE}{2\pi}
\big[ \;
G^{+-}_{Le\uparrow,Le\uparrow}(E)G^{-+}_{1e\uparrow,1e\uparrow}(E+\hbar\omega)
-G^{+-}_{1e\uparrow,Le\uparrow}(E)G^{-+}_{1e\uparrow,Le\uparrow}(E+\hbar\omega)\nonumber\\
-G^{+-}_{Le\uparrow,1e\uparrow}(E)G^{-+}_{Le\uparrow,1e\uparrow}(E+\hbar\omega)
+G^{+-}_{1e\uparrow,1e\uparrow}(E)G^{-+}_{Le\uparrow,Le\uparrow}(E+\hbar\omega)\big].
\end{eqnarray}
\end{widetext}
This approach is equivalent to the one presented by Chevallier et al. [\onlinecite{Chevallier}], but it is more transparent; one needs to determine only the Keldysh Green function with the electrodes, as that one (\ref{eq:NEGF}).

Notice that in Eq.(\ref{eq:greentot}) the Keldysh Green function matrix  is decoupled, which  means that there are two uncorrelated Andreev scattering channels: (e$\uparrow$,h$\downarrow$) and (h$\downarrow$,e$\uparrow$); and the corresponding currents are uncorrelated as well - similarly to the case in the paramagnetic system where the currents for electrons with the spin $\uparrow$ and $\downarrow$ are uncorrelated for a noninteracting case \cite{Bulka2000}. Thus, the cross correlations from two different subspaces are: $S_{\alpha e\uparrow,\alpha e\downarrow}= S_{\alpha h\uparrow,\alpha h\downarrow}=0$, and  $S_{Le\sigma,Re\sigma}= S_{Lh\sigma,Rh\sigma}= 0 $. This means that the direct Andreev reflections (DARs) as well as the single particle transfers (EC) are absent, and  transport is only due to perfect cross Andreev reflections (CARs). The situation would be different for a large transport window, when high energy processes (with  Cooper pair transfers through a single QD) should be taken into account.
Further, we will consider the (e$\uparrow$,h$\downarrow$) channel only, and to simplify the notation we will omit the spin indices  $\sigma=\uparrow,\downarrow$, which are related to the electron ($e$) and the hole ($h$), respectively.

Using the Green function, Eq.(\ref{eq:NEGF}), one can derive the current correlation function
\begin{align}\label{S-gen}
S_{\alpha,\alpha'}(\omega)= \frac{2e^2}{\hbar}\sum_{\nu,\nu'=Le,Rh}\int_{-\infty}^{\infty} \frac{dE}{2\pi}\;\mathscr{S}_{\alpha,\alpha'}^{\nu,\nu'}(E,E+\hbar\omega)\times
\nonumber\\ [1-f_{\nu}(E)]f_{\nu'}(E+\hbar\omega)\;,
\end{align}
where its the density elements are
\begin{align}
\label{sLLLL}
\mathscr{S}_{Le,Le}^{LeLe}(E,E&+\hbar\omega)=
1-r(E) r^*(E+\hbar\omega)-\nonumber\\
&r^*(E)r(E+\hbar\omega) +\mathcal{R}(E)\mathcal{R}(E+\hbar\omega),\\
\label{sLLRR}
\mathscr{S}_{Le,Le}^{RhRh}(E,E&+\hbar\omega)=
\mathcal{T}_{Aeh}(E)\mathcal{T}_{Aeh}(E+\hbar\omega),\\
\label{sLLRL}
\mathscr{S}_{Le,Le}^{RhLe}(E,E&+\hbar\omega)=
\mathcal{T}_{Aeh}(E) \mathcal{R}_{Aeh}(E+\hbar\omega), \\
\label{sLLLR}
\mathscr{S}_{Le,Le}^{LeRh}(E,E&+\hbar\omega)=\mathcal{R}_{Aeh}(E)
\mathcal{T}_{Aeh}(E+\hbar\omega)
\end{align}
for the auto-correlation function and
\begin{align}
\label{sLRLL}
\mathscr{S}_{Le,Rh}^{LeLe}(E,E+\hbar\omega)=&
t^*(E)t(E+\hbar\omega)
\nonumber\\
&[1 - r_{LL}^{*}(E+\hbar\omega)r_{LL}(E)],\\
\label{sLRRR}
\mathscr{S}_{Le,Rh}^{RhRh}(E,E+\hbar\omega)=&
t(E)t^{*}(E+\hbar\omega) \nonumber\\
&[1 - r_{RR}^*(E)r_{RR}(E+\hbar\omega) ] ,\\
\label{sLRLR}
\mathscr{S}_{Le,Rh}^{LeRh}(E,E+\hbar\omega)=&\mathscr{S}_{Le,Rh}^{RLeh}(E,E+\hbar\omega)=
\nonumber\\ &t^*(E)r_{LL}(E)t(E+\hbar\omega)r^*_{LL}(E+\hbar\omega)
\end{align}
for the cross-correlation function, respectively.  One can check that the formulas for $S_{\alpha,\alpha'}(\omega)$ have the same structure as those ones derived by means of the scattering matrix approach for the single quantum dot (1QD) attached to two normal electrodes~\cite{Buttiker1992a,Buttiker1992b,Dmytruk2016}. However, in our case the scattering matrix is different and its elements are
 \begin{align}\label{eq:t}
t(E)=\frac{4 \imath \sqrt{\gamma_L \gamma_R \gamma_{S}^2}}{4(z_{1e}-\imath\gamma_L) (z_{2h}-\imath\gamma_R)-\gamma_{S}^2},\\\label{eq:rll}
r_{LL}(E)=\frac{4 (z_{1e}+\imath \gamma_L)(z_{2h}-\imath\gamma_R)-\gamma_S^2}{4(z_{1e}-\imath\gamma_L) (z_{2h}-\imath\gamma_R)-\gamma_{S}^2},\\\label{eq:rrr}
r_{RR}(E)=\frac{4 (z_{1e}+\imath \gamma_L)(z_{2h}-\imath\gamma_R)-\gamma_S^2}{4(z_{1e}+\imath\gamma_L) (z_{2h}+\imath\gamma_R)-\gamma_{S}^2}.
\end{align}
and $\mathcal{T}_{Aeh}(E)=t(E)t^*(E)$, $\mathcal{R}_{Aeh}(E)=r_{LL}(E)r_{LL}^*(E)=r_{RR}(E)r_{RR}^*(E)=1-\mathcal{T}_{Aeh}(E)$.

\section{Results}
\label{results}

The preceding section has given the general analytical formulas for the current and their correlation. Let us now consider the results for special cases in detail.

\subsection{Zero frequency noise}
\label{zero-frequency}

First we present the zero-frequency current correlation function
\begin{align}\label{eq:slerh0}
&S_{Le,Le}(0)=S_{Le,Rh}(0)=
\nonumber\\
&\frac{2e^2}{\hbar}\int_{-\infty}^{\infty} \frac{dE}{2\pi}\;
\big\{\mathcal{T}_{Aeh}^2(E)[f_{Le}(1- f_{Le})+f_{Rh}(1- f_{Rh})]\nonumber\\  &+\mathcal{T}_{Aeh}(E)[1-\mathcal{T}_{Aeh}(E)][f_{Le}(1- f_{Rh})+f_{Rh}(1- f_{Le})]\big\}.
\end{align}
 The auto and cross correlations are equal due to charge current conservation. Moreover, this means a lack of direct Andreev reflections (DAR) and coherent cross Andreev reflections (CAR) are only responsible for transport through the system -- there is a perfect entanglement of Cooper pairs. The formula (\ref{eq:slerh0}) is similar to the noise power in a two-terminal conductor~\cite{Khlus1987,Lesovik1989,Buttiker1992a,Buttiker1992b, BlanterButtiker}, as well as in a hybrid structure in the presence of the electron-hole Andreev scattering~\cite{Anantram1996,Martin1996} (see also the review \cite{Lesovik2011,BlanterButtiker}). Note that the sign of the cross correlation function is positive because we consider the correlations between the electron and hole currents~\cite{Anantram1996,Martin1996}.  This in contrast to the case in the normal metallic Y splitter, where the cross correlation function corresponds to the electron-electron current correlations and its sign is negative~\cite{BlanterButtiker,Bulka2008}. Our calculations give an exact analytical expression (\ref{eq:TAeh})-(\ref{eq:TAhe}) for the transmission probabilities  $\mathcal{T}_{Aeh}$ and $\mathcal{T}_{Ahe}$.

In the linear response limit ($eV\rightarrow 0$) one can easily find the conductance and the spectral functions for the current correlations~\cite{Lesovik1999,Torres2001}.
 Let us consider the large bias voltage regime for the splitter bias configuration (with the chemical potentials $\mu_L=\mu_R=eV$ in both normal electrodes and $\mu_S=0$ in the superconductor), i.e.  for $f_{Le}=f_{Re}=0$ and $f_{Lh}=f_{Rh}=1$. In this case transport is unidirectional through both Andreev states, electrons are transmitted to the normal electrodes and holes in the opposite direction. There is no backscattering. The current can be determined as
\begin{align}
I_{Le}=\frac{e}{\hbar}\int_{-\infty}^{\infty}\frac{dE}{2\pi}\mathcal{T}_{Aeh}(E)
= \frac{e}{\hbar}\frac{\gamma \gamma_{S}^2}{ (\epsilon_1 + \epsilon_2)^2 + 4\gamma^2+ \gamma_{S}^2}
.
 \end{align}
 We consider only the single channel transport and therefore, the coefficient $e/\hbar$ appears.
The zero frequency cross correlation function is expressed as
\begin{align}
S_{Le,Rh}(0)=\frac{2e^2}{\hbar}\int_{-\infty}^{\infty}\frac{dE}{2\pi}\mathcal{T}_{Aeh}(E)[1- \mathcal{T}_{Aeh}(E)] \nonumber\\
=2eI_{Le}-\frac{2e^2}{\hbar}
\frac{ \gamma \gamma_{S}^4 [(\epsilon_1 +
      \epsilon_2)^2  + 20\gamma^2  + \gamma_{S}^2]}{2[ (\epsilon_1 + \epsilon_2)^2 + 4 \gamma^2+ \gamma_{S}^2]^3}
\;.\label{eq:slerhi0}
\end{align}
Here, to simplify the presentation we have assumed the symmetric coupling $\gamma_L=\gamma_R=\gamma$. The integration was performed using the residue theorem with the poles in the upper half plane: $ E_{\pm}=\epsilon_{\pm}+\imath \gamma$,
where $ \epsilon_{\pm}=\{ (\epsilon_1-\epsilon_2) \pm \delta\}/2$ denotes the position of ABS;  $\delta=\epsilon_+-\epsilon_-=\sqrt{ (\epsilon_1+\epsilon_2)^2 +\gamma_S^2}$ is the separation between them.
One can also write the Fano factor
\begin{eqnarray}
&F_{Le,Rh}\equiv \frac{S_{Le,Rh}(0)}{2|eI_{Le}|}=1-
\frac{\gamma_{S}^2 [(\epsilon_1 +
      \epsilon_2)^2  +  20\gamma + \gamma_{S}^2]}{2[ (\epsilon_1 + \epsilon_2)^2+  4\gamma^2+ \gamma_{S}^2]^2}
\;.
\nonumber\\
\end{eqnarray}
It is seen that its minimal value  $\min[F_{Le,Rh}]=7/32$ at $\gamma=\sqrt{3/20}\gamma_{S}$ and $\epsilon_1=-\epsilon_2$.

All analytical results were verified for $\omega=0$ by the full counting statistics (FCS) procedure~\cite{Dong,Gogolin2006}, when counting fields $\bm{\lambda}$ were introduced to the transfer Hamiltonian (\ref{eq:transferH}), to the Keldysh Green function (\ref{eq:NEGF}) and a proper adiabatic potential $\mathcal{U}(\bm{\lambda})$ was derived and used to calculate currents and their variances.

\subsection{Frequency dependence of current correlations}
\label{frequency}

\subsubsection{Large bias}

\begin{figure}
\includegraphics[width=1.0\linewidth,clip]{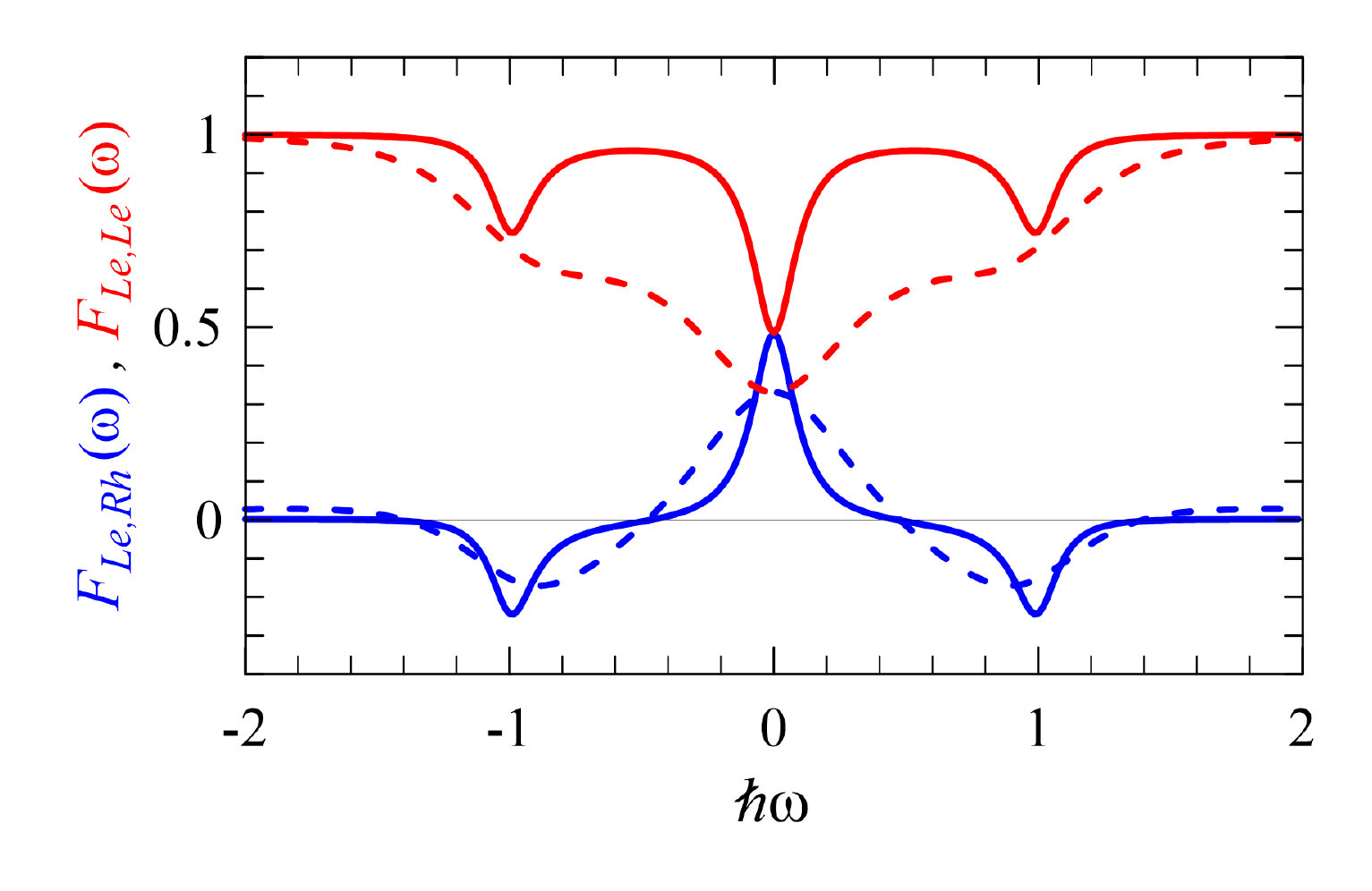}
\caption{Frequency dependence of the Fano factor: $F_{Le,Rh}$ (blue) and  $F_{Le,Le}$ (red) for  the cross- and auto-current correlation function, calculated  for the large voltage limit.
The solid curves represent the weak coupling $\gamma=0.05$, while the dashed curves are for the strong coupling $\gamma=0.2$. The other parameters are: $\gamma_S=1$, $\epsilon_1=\epsilon_2=0$, for which the separation between  ABS is $\delta=1$ and it is taken as the unit for the energy scale.}
\end{figure}

In the large bias voltage regime both ABS are in the voltage window and transport is unidirectional, backscattering processes are absent.
For the splitter voltage configuration the Fermi distribution functions $f_{Le}(E)=f_{Re}(E)=0$ and $f_{Lh}(E)=f_{Rh}(E)=1$. Moreover we assume moderate frequencies, $\hbar|\omega|\ll e|V|$, for which $f_{Le}(E+\hbar\omega)=f_{Re}(E+\hbar\omega)=0$ and $f_{Lh}(E+\hbar\omega)=f_{Rh}(E+\hbar\omega)=1$. In this case the current correlation function is expressed only by the shot noise term
\begin{align}\label{eq:slerhw}
S_{Le,Rh}(\omega)= \frac{2e^2}{\hbar}\int_{-\infty}^{\infty} \frac{dE}{2\pi}\;
\mathscr{S}_{Le,Rh}^{LeRh}(E,E+\hbar\omega)
\end{align}
with the density function $\mathscr{S}_{Le,Rh}^{LeRh}$, given by Eq.(\ref{sLRLR}).
For the symmetric coupling one can derive (using the residuum theorem)
\begin{align}\label{eq:slerhws}
&S_{Le,Rh}(\omega)
=\frac{2e^2}{\hbar} \frac{\gamma \gamma_S^2}{ \delta^2 + 4\gamma^2}\bigg[
\frac{2\gamma^2(2 \delta^2 - \gamma_S^2)}{\delta^2(4\gamma^2 + \hbar^2\omega^2)} \nonumber\\
&- \frac{\gamma^2\gamma_S^2}{ \delta^2(\delta^2 + 4\gamma^2)}\frac{ \delta^2 - 4\gamma^2 +2 \delta (\delta - \hbar\omega)}{(\delta - \hbar\omega)^2 + 4\gamma^2} \nonumber\\
&- \frac{\gamma^2\gamma_S^2}{ \delta^2(\delta^2 + 4\gamma^2)} \frac{ \delta^2 -4 \gamma^2 + 2 \delta (\delta + \hbar\omega)}{(\delta + \hbar\omega)^2 +4 \gamma^2}\bigg].
\end{align}
We have performed the spectral decomposition, which allows insight into local charge fluctuations and find their characteristic frequencies related to relaxation  processes~\cite{Bulka1999,Bulka2000}. Here, the method shows the interplay of the currents flowing through the higher (+) and lower (-) ABS and their contribution to the correlation function (see Appendix \ref{appB} for details). The first term is positive and represents the correlation function, $S_{Le,Rh}^{++}$ and $S_{Le,Rh}^{--}$, for the currents flowing through the same ABS. The other two terms correspond to inter-level current correlations, $S_{Le,Rh}^{-+}$ and $S_{Le,Rh}^{+-}$, respectively. These
functions are negative and are related to emission and absorption of energy $\hbar\omega$  by the system.
For $\omega=0$ the result obviously coincides with Eq.(\ref{eq:slerhi0}). Notice that $S_{Le,Rh}(\omega)=S_{Lh,Re}(\omega)$  and they depend only on $|\epsilon_1+\epsilon_2|$, which means that fluctuations on both Andreev states are equivalent.

In a similar way we derive shot noise in the current auto-correlation function:
\begin{align}\label{eq:slele1}
&S_{Le,Le}(\omega)= \frac{2e^2}{\hbar}\frac{ \gamma \gamma_S^2}{\delta^2 +4 \gamma^2}\bigg[1-\frac{2\gamma^2 \gamma_S^2}{\delta^2(4\gamma^2+
 \hbar^2\omega^2)}\nonumber\\
 &- \frac{\gamma^2\gamma_S^2}{ \delta^2(\delta^2 + 4\gamma^2)}\frac{ \delta^2 - 4\gamma^2 +2 \delta (\delta - \hbar\omega)}{(\delta - \hbar\omega)^2 + 4\gamma^2} \nonumber\\
&- \frac{\gamma^2\gamma_S^2}{ \delta^2(\delta^2 + 4\gamma^2)} \frac{ \delta^2 -4 \gamma^2 + 2 \delta (\delta + \hbar\omega)}{(\delta + \hbar\omega)^2 +4 \gamma^2}\bigg].
\end{align}
 The first term of $S_{Le,Le}(\omega)$ corresponds to the Schottky noise, which is frequency independent and describes uncorrelated transfers of particles with a Poissonian distribution function of time intervals between transfer events~\cite{BlanterButtiker}. The Pauli principle and the symmetry of electron wave functions result in the negative current correlations, which is also a signature of anti-bunching of the transfer events,  as occurs in our case for the second term in Eq.(\ref{eq:slele1}).

In contrast the first term in   $S_{Le,Rh}(\omega)$, Eq.(\ref{eq:slerhws}), is positive, which means that the electron and hole transfers, in the Andreev scattering, are bunched. The two last terms are the same in $S_{Le,Rh}(\omega)$  and $S_{Le,Le}(\omega)$. They are related to correlation of the currents flowing through the lower and the upper ABSs, which is  maximal at $\hbar\omega=\pm\delta$ when the resonant peaks of $\mathcal{T}_{Aeh}(E)$ and $\mathcal{T}_{Aeh}(E\pm\delta)$ overlap with each other. The sign of these correlations
is negative which means that inter-level transfers are anti-bunched. The similar situation can occur in the multilevel quantum dot system, when a transmission coefficient has many resonant peaks, which can lead to negative and sometimes positive current correlations.
\begin{figure}
\includegraphics[width=.85\linewidth,clip]{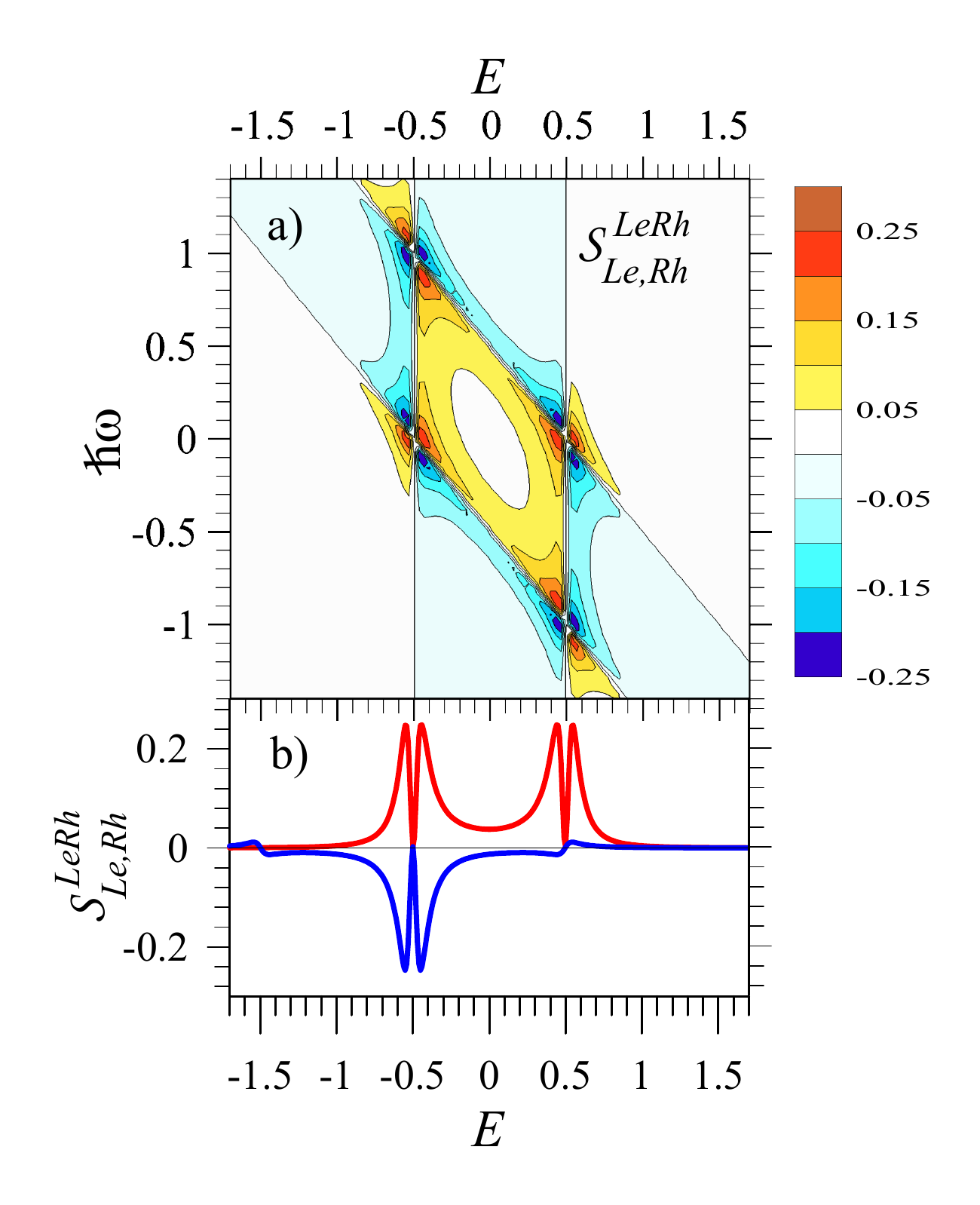}
\includegraphics[width=.85\linewidth,clip]{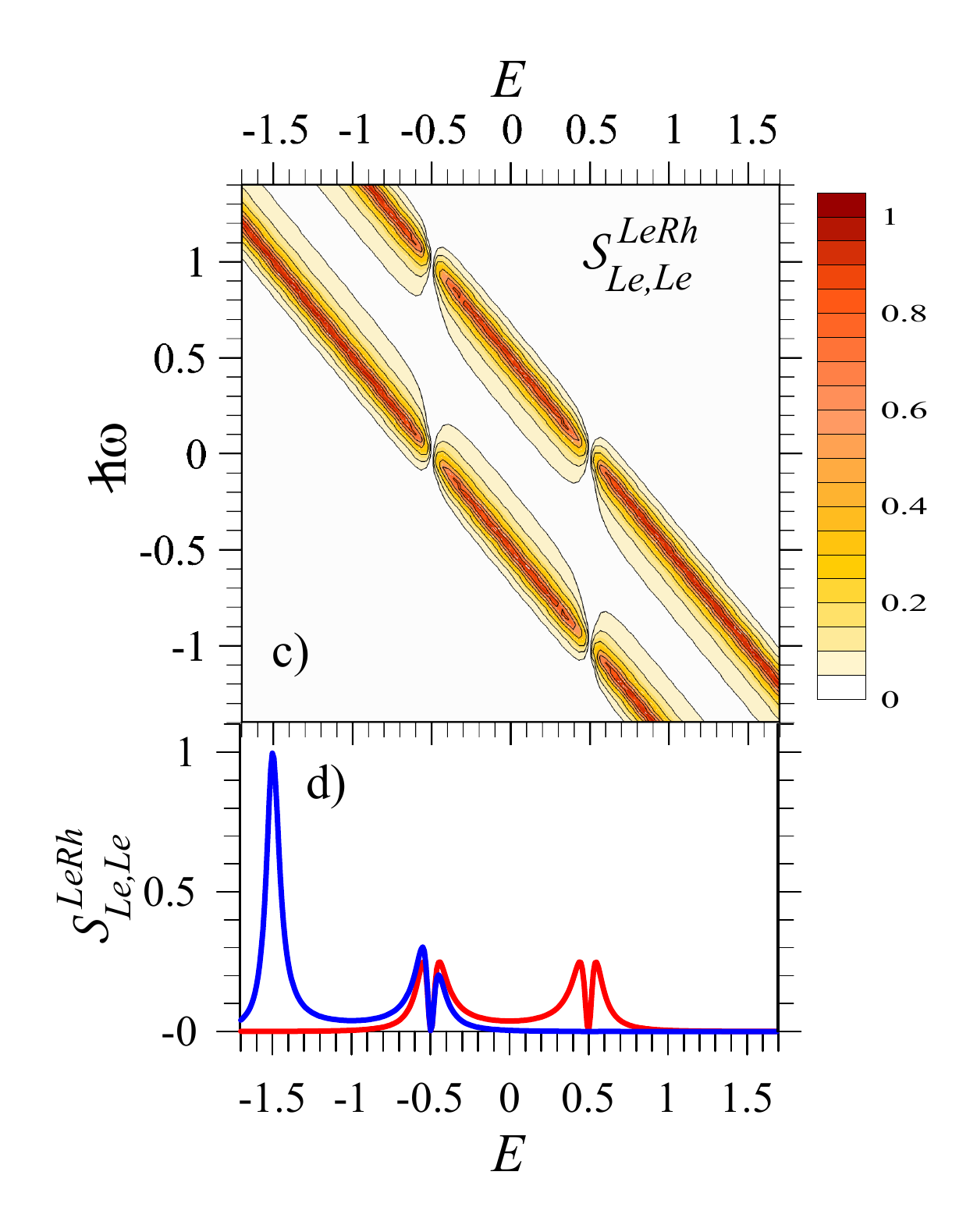}
\caption{Map of the density function:  $\mathcal{S}^{LeRh}_{Le,Rh}(E,E+\hbar\omega)$ -- (a) and $\mathcal{S}^{LeRh}_{Le,Le}(E,E+\hbar\omega)$ -- (c), plotted vs the energy $E$ and the frequency $\hbar\omega$.  Panels (b) and (d), present the plots of the density functions vs $E$ at $\hbar\omega=0$ and $\hbar\omega=1$, the red and the blue curves, respectively.
  The plots are performed for the weak coupling $\gamma_L=\gamma_R=0.05$ and the other parameter are the same as in Fig.2.
}\label{fig3}
\end{figure}

The frequency dependence of the Fano factor $F_{Le,Rh}(\omega)\equiv S_{Le,Rh}(\omega)/2eI_{Le}$ and $F_{Le,Le}(\omega)\equiv S_{Le,Rh}(\omega)/2eI_{Le}$ for the cross and auto-correlations
are shown in Fig.2. In this case normalization is given by  $2eI_{Le}=(2e^2/\hbar)\gamma\gamma_S^2/(\delta^2+4\gamma^2)$.

A similar frequency dependence was found by Droste et al.~\cite{Droste}
for the correlation function in the two terminal hybrid system, with one quantum dot (N-QD-S), which was derived by means of a real-time diagrammatic perturbation expansion in the tunnel-coupling to the normal electrode. The authors argue~\cite{Droste}, that the dips
arise from a coherent destructive interference between ABS. However, our studies show that the origin of the dips is different. The transport in both the models is coherent, but there is no evidence for destructive interference in the transmission coefficient $\mathcal{T}_{Aeh}(E)$ and in the differential conductance. The spectral decomposition analysis shows that the side dips are due to the negative current correlations related with photon assisted electron transfers through the different ABS. Such structure of the noise power spectrum can occur also for coherent transport through
multilevel quantum dots~\cite{Entin,Rothstein,Marcos}.
In particular,  Ref.~\cite{BulkaJMMM} showed that electron-hole recombination processes can result in resonant inter-level current correlations with emission/absorption of energy in the system of two quantum dots in a T-geometry coupled to the normal metallic electrodes.

Our results are in contrast to those in Ref.\cite{Michalek2021} where the noise power spectrum in 2QD-CPS was determined in the sequential tunneling regime based on a diagonalized master equation (DME). The method gives reliably the voltage dependence of the current, which reflects the internal spectrum of the proximized 2QDs with ABSs.
The spectral decomposition analysis showed contributions of current correlations through various ABSs  and the interplay between different local  charge relaxation processes.
However, the DME method neglects coherent electron-hole recombination processes, and therefore, any extraordinary side
dips have not been found in the noise power spectrum.

To have a better insight into the role of quantum interference we analyze the density functions $\mathscr{S}_{Le,Rh}^{\nu\nu'}$ and $\mathscr{S}_{Le,Le}^{\nu\nu'}$ for the cross- and the auto-current correlation, respectively. In the shot noise regime the relevant components are:
\begin{align}
\mathscr{S}^{LeRh}_{Le,Rh}=t^*(E)r_{LL}(E)t(E+\hbar\omega)r^*_{LL}(E+\hbar\omega)
\end{align}
and
\begin{align}
\mathscr{S}^{LeRh}_{Le,Le}=\mathcal{T}_{Aeh}(E+\hbar\omega)
[1-\mathcal{T}_{Aeh}(E)].
\end{align}
Figure 3 presents their plots as maps in the energy and frequency space, $(E,\hbar\omega)$. The function $\mathscr{S}^{LeRh}_{Le,Le}$ has two peaks shifted in energy by $\delta$, which is related to the spectrum of ABS seen in $\mathcal{T}_{Aeh}(E+\hbar\omega)$ and $[1-\mathcal{T}_{Aeh}(E)]$.
The density function $\mathscr{S}^{LeRh}_{Le,Rh}$ is more interesting, which can be positive or negative. For example, it is positive for $\hbar\omega=0$ and equal to $\mathcal{T}_{Aeh}(1-\mathcal{T}_{Aeh})$, while it is negative for whole energy range at  $\hbar\omega=\pm1$ - see Fig.3b.

\subsubsection{Moderate bias voltage}

For $V\rightarrow \infty$ we have been able to derived the analytical formulas, but for a finite voltage one
should use the general formula for the correlation functions, Eq.(\ref{S-gen}), taking into account all components of  the density functions, Eqs.(\ref{sLLLL})-(\ref{sLRLR}). The integrals have been calculated numerically for the temperature $T=0$ and the results for the frequency dependent current correlation functions are presented in Fig.4. Comparing with Fig.2 (for $V\rightarrow \infty$) one sees that the plots vanish for the positive $\hbar\omega$ when the voltage $eV$ becomes smaller. As expected with decreasing voltage less noise power is emitted by the system. On the other hand the other components, those related to thermal fluctuations on the same junctions, become more relevant, and they result in an increase of the noise power spectrum, in particular for $\hbar\omega<0$.  The plots show the pronounced dips at $\hbar\omega=-1$ (in particular that  for $S_{Le,Rh}$), which are related to negative inter-level current correlations with energy absorption.

\begin{figure}
\includegraphics[width=.8\linewidth,clip]{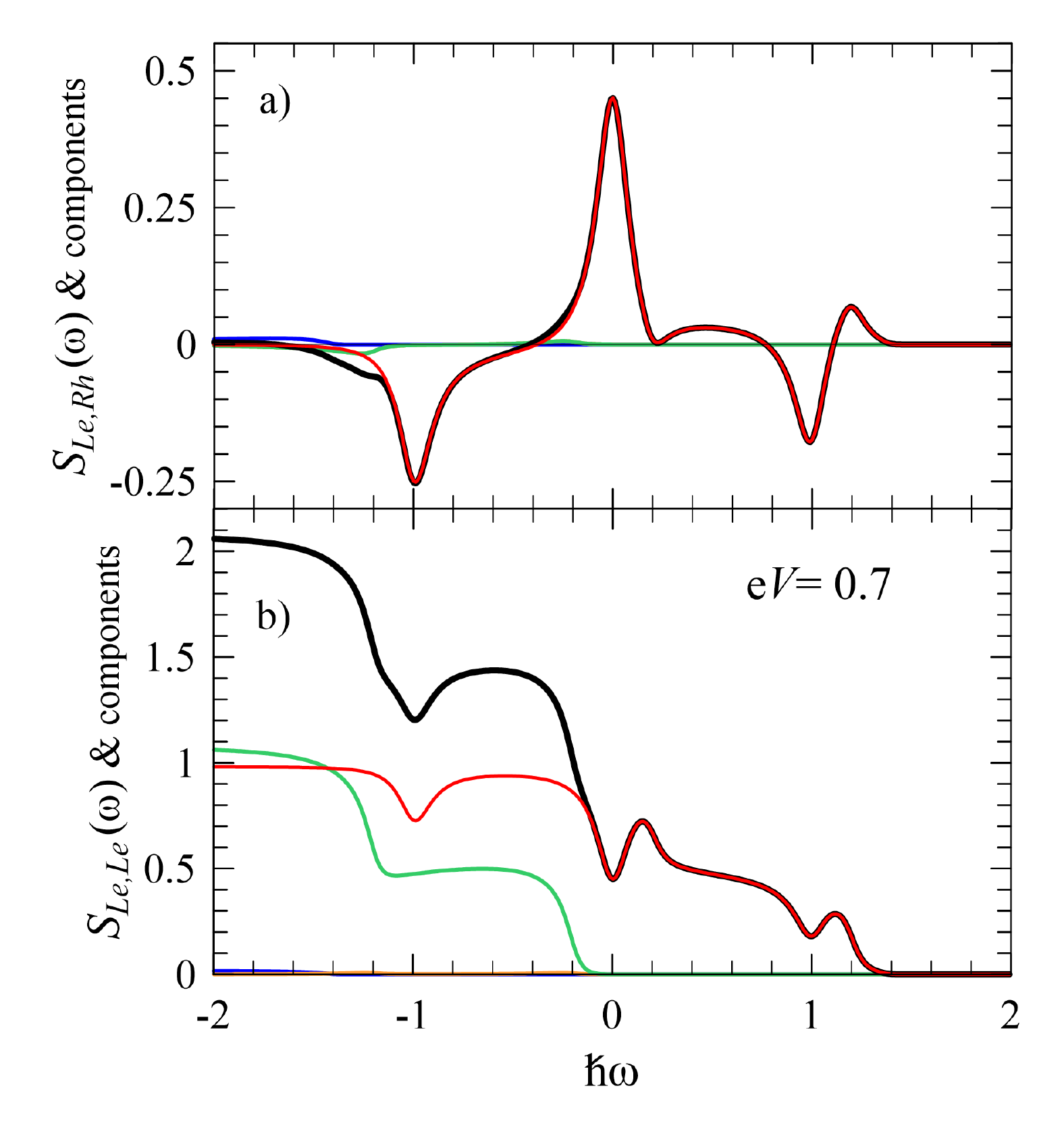}
\includegraphics[width=.8\linewidth,clip]{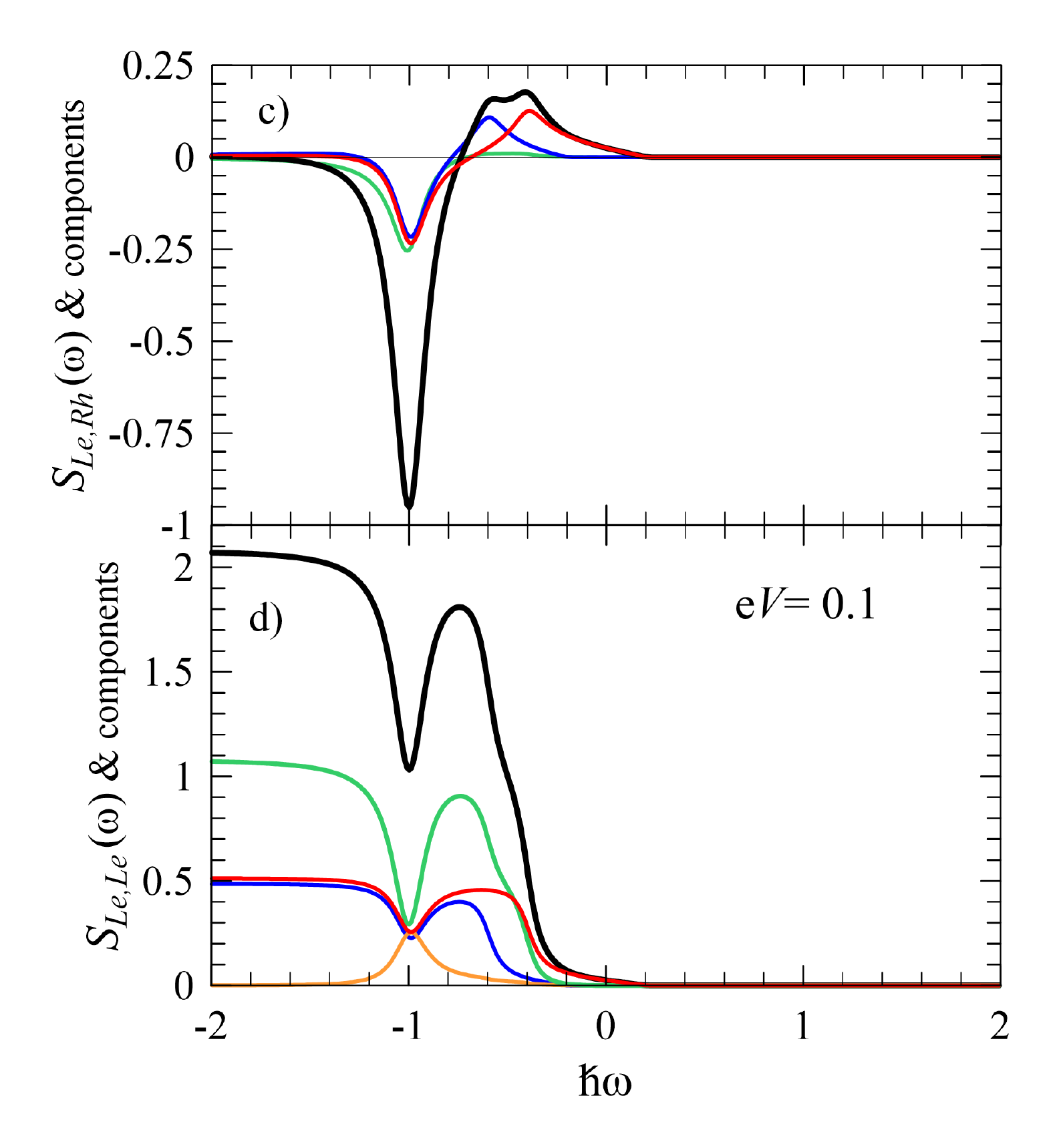}
\caption{Frequency dependence of the real part of the cross- and the auto-correlation function (black thick curves) with their components: $S_{Le,Rh}^{LeRh}$ and $S_{Le,Le}^{LeRh}$ (red),  $S_{Le,Rh}^{RhLe}$ and $S_{Le,Le}^{RhLe}$ (blue),  $S_{Le,Rh}^{LeLe}=S_{Le,Rh}^{RhRh}$ and $S_{Le,Le}^{LeLe}$ (green), $S_{Le,Le}^{RhRh}$ (orange), for $eV=0.7$  [(a) and (b)] and $eV=0.1$ [(c) and (d)]. The plots are calculated at temperature $T=0$ and normalized to $2e\max[I_{Le}]=(2e^2/\hbar)\gamma\gamma_S^2/(\delta^2+4\gamma^2)$.
The other parameter are the same as in Fig.3.
}\label{fig4}
\end{figure}

\subsubsection{Linear response limit}

\begin{figure}
\includegraphics[width=.9\linewidth,clip]{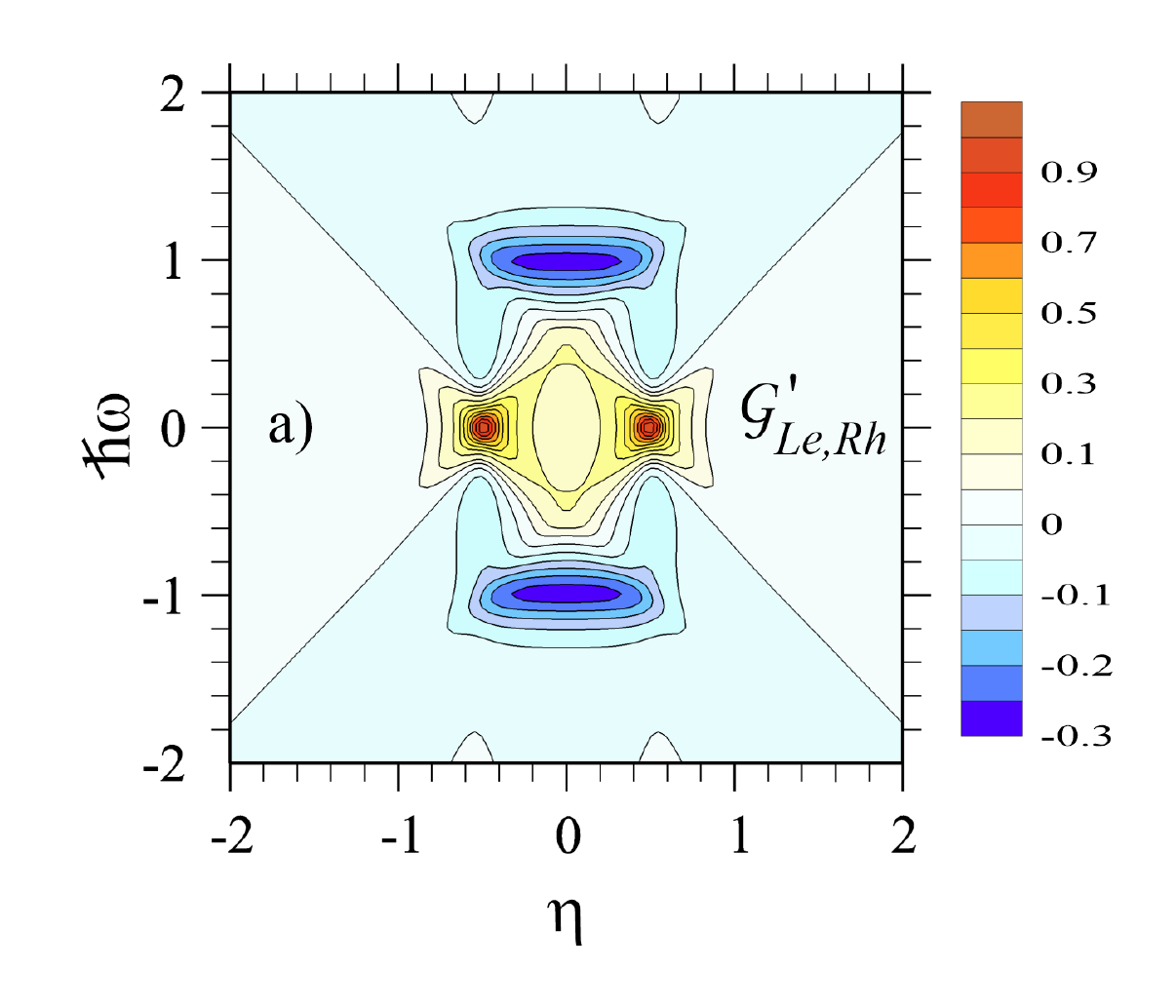}
\includegraphics[width=.9\linewidth,clip]{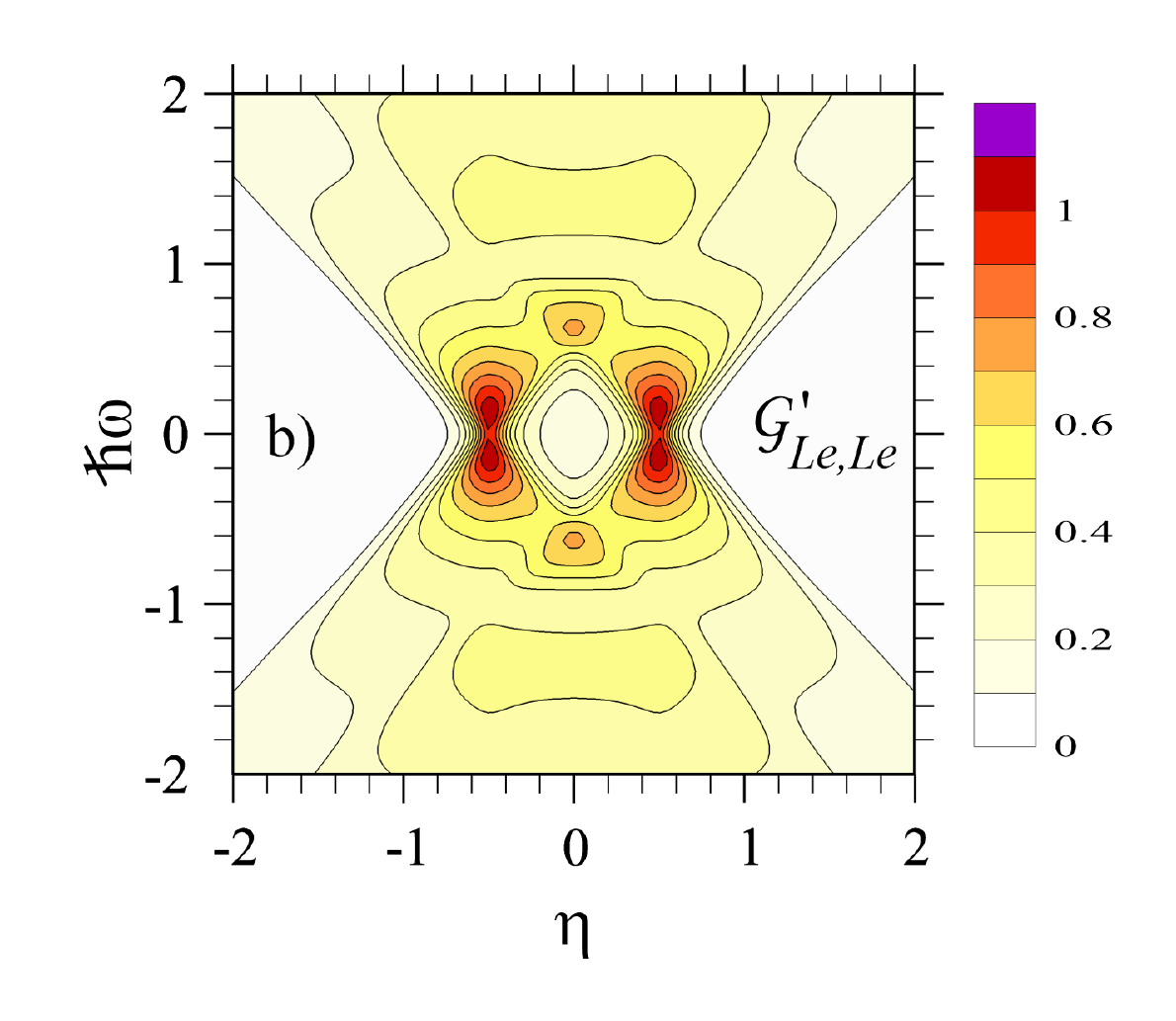}
\caption{ Map of the real part of the admittance (a) $\mathcal{G}'_{Le,Rh}$ and (b) $\mathcal{G}'_{Le,Le}$ (in units $2e^2/h$) plotted in the space $(\eta,\hbar\omega)$, where $\eta=(\epsilon_1-\epsilon_2)/2$.  We take $\epsilon\equiv(\epsilon_1+\epsilon_2)/2=0$, $E_F=0$, temperature $T=0$,  and the other parameters are the same as in Fig.3.  }\label{fig5}
\end{figure}

For a small bias voltage one can use the Kubo theory of linear response  and establish the relation
\begin{align}\label{admit}
S_{\alpha\alpha'}(\omega)-S_{\alpha'\alpha}(-\omega)
= -4\hbar\omega\mathcal{G}'_{\alpha\alpha'}(\omega)=
\nonumber\\
\frac{4e^2}{h}\int dE\; \mathscr{S}_{\alpha\alpha'}^{\text{eq}}(E,E+\hbar\omega) [f(E)-f(E+\hbar\omega)]
\;,
\end{align}
between the noise power spectrum and the real part of the admittance $\mathcal{G}_{\alpha\alpha'}(\omega)$ (responsible for dissipation)~\cite{Nazarov}. Here, we denoted
$\mathscr{S}_{\alpha\alpha'}^{\text{eq}}\equiv\sum_{\eta,\eta'=L,R} \mathscr{S}_{\alpha\alpha'}^{\eta\eta'}$, which can be explicitly expressed as
\begin{align}
&\mathscr{S}_{LL}^{\text{eq}}= 2-r_{LL}^*(E) r_{LL}(E+\hbar\omega)-r_{LL}(E) r_{LL}^*(E+\hbar\omega),\\
&\mathscr{S}_{LR}^{\text{eq}}= t(E) t^{*}(E+\hbar\omega) +t^*(E)t(E+\hbar\omega)
\end{align}
for the auto-and cross-correlations, respectively. [The coefficients $t(E)$ and $r_{LL}(E)$ are given by Eqs.(\ref{eq:t})-(\ref{eq:rll}).]
For $T=0$ one gets
\begin{align}\label{admit0}
 4\hbar|\omega|\mathcal{G}'_{\alpha\alpha'}(\omega)=
\frac{4e^2}{h}\int_{E_F}^{E_F+\hbar|\omega|} dE\; \mathscr{S}_{\alpha\alpha'}^{\text{eq}}(E,E+\hbar\omega)
\;,
\end{align}
where the Fermi energy $E_F=0$.
Notice that in this limit the noise is purely quantum:   $S_{\alpha\alpha'}(\omega)=-4\hbar\omega\mathcal{G}'_{\alpha\alpha'}(\omega)\theta(-\omega)$, where $\theta$ denotes the Heaviside step function.
Figure~\ref{fig5} presents the admittance,
$\mathcal{G}'_{Le,Rh}$ (a) and $\mathcal{G}'_{Le,Le}$ (b), plotted in the space $(\eta,\hbar\omega)$, where $\eta\equiv(\epsilon_1-\epsilon_2)/2$ and $\epsilon\equiv(\epsilon_1+\epsilon_2)/2=0$. Although for $\hbar\omega= 0$ one gets $\mathcal{G}'_{Le,Rh}(0)=\mathcal{G}'_{Le,Le}(0)= (2e^2/h)\mathcal{T}_{Aeh}(E_F)$, but for finite frequencies these admittances exhibit different interference patterns. It is seen that the cross admittance, $\mathcal{G}'_{Le,Rh}$, decreases with $\hbar|\omega|$ and becomes negative, while the auto admittance, $\mathcal{G}'_{Le,Le}$, is always positive and can be larger than $2e^2/h$ due to current fluctuations on the same tunnel junction. These results show that displacement currents play a different role in the cross and the auto correlations.

\section{Summary}

Let us summarize the main results. We analyzed  a Cooper pair splitter model, where two entangled electrons were transferred through the inter-dot singlet state on two proximized QDs into two normal electrodes. Since the derivations were confined to the subspace with the inter-dot singlet, where high energy processes were forbidden, we were able to show the separation of the cross  Andreev  reflections: for an electron-hole (e$\uparrow$,h$\downarrow$) and a hole-electron (h$\downarrow$,e$\uparrow$) scattering [see Eq.(\ref{eq:greentot})]. This means that the transfer of an electron with the spin $\uparrow$ from the first QD to the left electrode is coherently coupled (fully entangled) with the transfer of a hole with the spin $\downarrow$ from the right electrode to the second QD, and this process is independent from the (h$\downarrow$,e$\uparrow$) transfer. Consequently, the zero frequency cross and auto correlation functions are equal and positive, $S_{Le,Rh}(0)=S_{Le,Le}(0)>0$.

The other important result is the derivation of the analytical formulas for the frequency dependent  cross and auto-correlation functions,  $S_{Le,Rh}(\omega)$ and $S_{Le,Le}(\omega)$ in the large bias limit, Eq.(\ref{eq:slerhws}) and (\ref{eq:slele1}). Here, quantum coherence manifests itself in the negative inter-level current correlations as the dips at $\hbar\omega=\pm\delta$, when the system absorbs or emits the energy.
The intra-level current correlations are seen at the low frequency limit and they are responsible for the central dip in $S_{Le,Le}(\omega)$ as well as the central peak in $S_{Le,Rh}(\omega)$. These low-frequency features are well known in sequential and coherent transport through quantum dots  \cite{Buttiker1992a,Buttiker1992b, BlanterButtiker,Korotkov, Bulka2000}.

We also considered the admittance (derived by means of the linear response theory) for the cross- and the auto-correlation case, $\mathcal{G}_{Le,Le}$ and $\mathcal{G}_{Le,Rh}$, respectively. Although at $\omega=0$ they are equal and coincide with the conductance, for the finite $\omega$ they have different interference patterns.
Our results show that quantum interference manifests  itself in the second-order current correlation functions in a different way than Young's interference patterns in the first-order correlations.

We studied the noise power spectrum in the non-symmetrized form to get a direct relation with a quantum noise measurement. To verify our predictions one can use the experimental setup as in Ref.~\cite{Aguado,Clerk2002, Deblock2003,Xue2009, Billangeon2006,Billangeon2009, Basset}, with a quantum detector sensitive in a large frequency range to detect inter-level transitions between ABSs. This seems to be an experimental challenge.

\acknowledgments{
The research was financed by National Science Centre,
Poland, Project No. 2016/21/B/ST3/02160.}

\appendix
\section{\label{appA} Self-energies for N and S-electrode}

We use the equation of motion to derive the Green functions in the Keldysh and Nambu spaces. In this method important quantities are the self-energies which describe coupling of the system with the normal  and superconducting electrodes. For the normal electrode coupling one simply gets the self-energy
\begin{widetext}
\begin{eqnarray}\label{eq:normal}
\hat{\Sigma}_{\alpha i}=\frac{\imath\gamma_{\alpha i}}{2}
\left[\begin{array}{cccc}
2f_{\alpha e}-1&2f_{\alpha e}&0&0\\
2(f_{\alpha e}-1)&2f_{\alpha e}-1&0&0\\
0&0&2f_{\alpha h}-1&2f_{\alpha h}\\
0&0&2(f_{\alpha h}-1)&2f_{\alpha h}-1
\end{array}\right],
\end{eqnarray}
where the wide flat-band approximation is used, and $\gamma_{\alpha i}=\pi \rho_{\alpha} |t_{\alpha i}|^2$, $\rho_{\alpha}$ denotes the density of states in the electrode, $f_{\alpha e}=\{\exp[(E-\mu_{\alpha})/k_BT]+1\}^{-1}$ and  $f_{\alpha h}=\{\exp[(E+\mu_{\alpha})/k_BT]+1\}^{-1}$ are the distribution functions for electrons and holes, respectively. For the coupling of the S-electrode with the 1 and 2 QD the self-energy is expressed as (following [\onlinecite{Chevallier,Dong,Zazunov,Cuevas}]):
\begin{align}\label{eq:S-Sd}
&\hat{\Sigma}_{S}=\frac{\imath\gamma_{S}}{2}\times\nonumber\\
&\left[\begin{array}{cccc}
(2f_{Se}-1)\beta^{'}_S+\imath \beta^"_S& 2f_{Se}\beta^{'}_S&-[(2f_{Se}-1)\beta^{'}_S+\imath \beta^"_S]\Delta/E&-2f_{Se}\beta^{'}_S \Delta/E\\
2(f_{Se}-1)\beta^{'}_S&(2f_{Se}-1)\beta^{'}_S+\imath \beta^"_S&2f_{Se}\beta^{'}_S \Delta/E&[(2f_{Se}-1)\beta^{'}_S+\imath \beta^"_S]\Delta/E\\
 ((2f_{Sh}-1)\beta^{'}_S-\imath \beta^{"}_S)\Delta/E&
 2(f_{Sh}-1)\beta^{'}_S \Delta/E&(2f_{Sh}-1)\beta^{'}_S-\imath \beta^"_S&2f_{Sh}\beta^{'}_S\\
 -2(f_{Sh}-1)\beta^{'}_S \Delta/E&-[(2f_{Sh}-1)\beta^{'}_S-\imath \beta^"_S]\Delta/E &2(f_{Sh}-1)\beta^{'}_S&(2f_{Sh}-1)\beta^{'}_S-\imath \beta^"_S
\end{array}\right],\nonumber\\
\end{align}
\end{widetext}
where $\gamma_{S}=\pi \rho_S t_{L1}t_{2R}$, $\beta^{'}_S$ and $\beta^"_S$ are the real and the imaginary part of the function
\begin{eqnarray}
\beta_S=\frac{|E|\theta(|E|-\Delta)}{\sqrt{E^2-\Delta^2}}- \imath\frac{E\theta(\Delta-|E|)}{\sqrt{\Delta^2-E^2}}\;.
\end{eqnarray}
Notice that in the limit $\Delta \rightarrow 0$ one gets $\hat{\Sigma}_{S}$ as for the normal electrode, Eq.(\ref{eq:normal}).
For the sub-gap regime $|E|<\Delta$ the function $\beta_S=-\imath E/\sqrt{\Delta^2-E^2}$ and in the limit $\Delta \rightarrow \infty$  Eq.(\ref{eq:S-Sd}) reduces itself to Eq.(\ref{eq:S-S}) in Section \ref{model}.

Let us stress that in a similar way one can derive the self-energy $\hat{\Sigma}_{S1}$ and $\hat{\Sigma}_{S2}$ related to the intra-dot pairing. However, in our considerations the intra-dot Coulomb interactions are assumed to be large and these terms are neglected as they correspond to high energy tunneling processes.

\section{\label{appB} Spectral decomposition}

Let us present the spectral decomposition approach which allows to separate various contributions of currents flowing through ABSs and their components in noise power. Our derivations are performed in the large voltage regime, in which the current
\begin{eqnarray}
I_{Le} =-\frac{e}{\hbar}\int_{-\infty}^{\infty} \frac{dE}{2\pi} \mathcal{T}_{Aeh}(E)
\end{eqnarray}
and we want to express it as $I_{Le}=I_{Le}^+ + I_{Le}^-$, separating the current flowing through the upper ($+$) and the lower ($-$) ABSs. To this end the transmission coefficient $\mathcal{T}_{Aeh}$, Eq.(\ref{eq:TAeh}), is decomposed into partial fractions and the terms are grouped into those corresponding to transmission through the upper and the lower state:
 \begin{align}\label{T-spect}
 \mathcal{T}^{\pm}_{Aeh}(E)=& \mp\frac{\imath \gamma \gamma_S^2}{ 2 \delta(\delta^2+4\gamma^2)}\Big[ \frac{-2 \imath \gamma\pm\delta}{ E - \epsilon_{\pm} - \imath \gamma } - \frac{2 \imath \gamma\pm\delta}{ E - \epsilon_{\pm} + \imath \gamma }\Big]
 \nonumber\\
 =& \mp\frac{ 2\gamma^2 \gamma_S^2( E-\epsilon_{\pm} \mp\delta/2)}{\delta(4\gamma^2+\delta^2)[(E-\epsilon_{\pm})^2+\gamma^2]},
 \end{align}
where $ \epsilon_{\pm}$ denotes the position of the ABS. Thus, the current is $I_{Le}^+=I_{Le}^-=(e/\hbar)\gamma\gamma_S^2/[2(4\gamma^2+\delta)]$.
Here, we assume the symmetric coupling, $\gamma_L=\gamma_R=\gamma$.

Similarly, we perform spectral decomposition of the cross correlation function $S_{Le,Rh}^{LeRh}(\omega)$, where the corresponding density $\mathscr{S}_{Le,Rh}^{LeRh}(E,\omega)=t^*(E)r_{LL}(E)t(E+\hbar\omega)r^*_{LL}(E+\hbar\omega)= g(E)g^*(E+\hbar\omega)$, Eq.(\ref{sLRLR}). To this end the function $g(E)\equiv\imath t^*(E)r_{LL}(E)$ is decomposed into partial fractions and next, the terms are grouped for those for transfers through $\epsilon_+$ and $\epsilon_-$. The result is
\begin{align}
&g_{\pm}(E)=\pm\frac{\gamma\gamma_S}{2\delta(\delta^2+4\gamma^2)}\Big[
\frac{(\epsilon_1+\epsilon_2\pm\delta)(-2\imath\gamma\pm\delta)}{E- \epsilon_{\pm}-\imath\gamma}\nonumber\\
&\:\;\;-\frac{(\epsilon_1+\epsilon_2\mp\delta)(2\imath\gamma\pm\delta)}{E- \epsilon_{\pm}+\imath\gamma}\Big]
=\mp\frac{\gamma\gamma_S}{\delta(\delta^2+4\gamma^2)}\times\nonumber\\
&\frac{(E-\epsilon_{\pm})\delta^2 \pm 2\delta\gamma^2-2\imath(\epsilon_1+\epsilon_2) (E-\epsilon_{\pm}\mp\delta/2)} {(E-\epsilon_{\pm})^2+\gamma^2}.
\end{align}
Thus, the function
\begin{align}
\mathscr{S}_{Le,Rh}^{LeRh}(E,\omega)=\sum_{\nu,\nu'=\pm} \mathscr{S}_{Le,Rh}^{LeRh,\nu\nu'}(E,\omega),\end{align}
where its density $\mathscr{S}_{Le,Rh}^{LeRh,\nu\nu'}(E,\omega)= g_{\nu}(E)g_{\nu'}^{*}(E+\hbar\omega)$.

After integration one gets the spectral components of $S_{Le,Rh}^{LeRh}(\omega)$:
\begin{align}
&S_{Le,Rh}^{LeRh,++}(\omega)=\frac{2e^2}{\hbar}
\frac{\gamma^2\gamma_S^2[(2\delta^2-\gamma_S^2)\delta +\imath(\epsilon_1+\epsilon_2)\delta]} {\delta^2(4\gamma^2+\delta^2)(4\gamma^2 +\hbar^2\omega^2)}
,\\
\label{slerh+-}
&S_{Le,Rh}^{LeRh,+-}(\omega)=-\frac{2e^2}{\hbar}\frac{\gamma^3\gamma_S^4
[\delta^2-4\gamma^2+2\delta(\delta+\hbar\omega)]}{\delta^2(4\gamma^2+\delta^2)^2[4\gamma^2 +(\delta+\hbar\omega)^2]}
,\\
\label{slerh-+}
&S_{Le,Rh}^{LeRh,-+}(\omega)=-\frac{2e^2}{\hbar}\frac{\gamma^3\gamma_S^4
[\delta^2-4\gamma^2+2\delta(\delta-\hbar\omega)]}{\delta^2(4\gamma^2+\delta^2)^2[4\gamma^2 +(\delta-\hbar\omega)^2]},\\
&S_{Le,Rh}^{LeRh,--}(\omega)=\frac{2e^2}{\hbar}\frac{\gamma^2\gamma_S^2[(2\delta^2-\gamma_S^2)\delta -\imath(\epsilon_1+\epsilon_2)\delta]} {\delta^2(4\gamma^2+\delta^2)(4\gamma^2 +\hbar^2\omega^2)}.
\end{align}

Let us perform spectral decomposition of $S_{Le,Le}^{LeRh}(\omega)$, for which the density is expressed as $\mathscr{S}_{Le,Le}^{LeRh}(E,\omega)=\mathcal{T}_{Aeh}(E+\hbar\omega)- \mathcal{T}_{Aeh}(E+\hbar\omega) \mathcal{T}_{Aeh}(E)$. Since $\mathcal{T}_{Aeh}(E+\hbar\omega)$ gives the frequency independent Schottky term in the noise power, we focus on the second term which can be written as
$\mathcal{T}_{Aeh}(E+\hbar\omega) \mathcal{T}_{Aeh}(E)= \sum_{\nu,\nu'=\pm}\mathcal{T}^{\nu}_{Aeh}(E+\hbar\omega) \mathcal{T}^{\nu'}_{Aeh}(E)$. Using Eq.(\ref{T-spect}) and integrating one can find the spectral components of the frequency dependent part of $S_{Le,Le}^{LeRh}(\omega)$:

\begin{align}
\frac{2e^2}{\hbar}\bigg\{&-\frac{2\gamma^3 \gamma_S^4}{\delta^2(4\gamma^2+ \hbar^2\omega^2)}\nonumber\\
&- \frac{\gamma^3\gamma_S^4[ \delta^2 - 4\gamma^2 +2 \delta (\delta + \hbar\omega)]}{ \delta^2(\delta^2 + 4\gamma^2)^2[4\gamma^2+(\delta + \hbar\omega)^2]} \nonumber\\
&- \frac{\gamma^3\gamma_S^4[ \delta^2 - 4\gamma^2 +2 \delta (\delta - \hbar\omega)]}{ \delta^2(\delta^2 + 4\gamma^2)^2[ 4\gamma^2+(\delta - \hbar\omega)^2]}\bigg\}.
\end{align}
Here, the first term comes from the $(++)$ and $(--)$ component, while the second and third term comes from the inter-level $(+-)$ and $(-+)$ correlation, and they are equal to $S_{Le,Rh}^{LeRh,+-}(\omega)$ and $S_{Le,Rh}^{LeRh,-+}(\omega)$, Eqs.(\ref{slerh+-}) and (\ref{slerh-+}), respectively.

%
\bibliographystyle{apsrev4-2}
\bibliography{2qd-splitter-b}

\end{document}